\documentclass[traditabstract]{aa} 
\usepackage{graphicx}
\usepackage{natbib}
\usepackage{txfonts}
\begin{document}
 \title{Updated parameters for the transiting exoplanet WASP-3b using RISE, a new fast camera for the Liverpool Telescope}
 \titlerunning{Updated parameters for the transiting exoplanet WASP-3b using RISE}
 
   \author{N. P. Gibson
          \inst{1}\fnmsep\thanks{\email{ngibson07@qub.ac.uk}}\and
          D. Pollacco\inst{1}\and
         E. K. Simpson\inst{1}\and
         Y. C. Joshi\inst{1}\and
         I. Todd\inst{1}\and
         C. Benn\inst{2}\and
         D. Christian\inst{1}\and
         M. Hrudkov\'{a}\inst{3}\and
         F. P. Keenan\inst{1}\and
	J. Meaburn\inst{4}\and
	I. Skillen\inst{2}\and
	I. A. Steele\inst{5}.
          }

   \institute{
   	Astrophysics Research Centre, School of Mathematics \&\ Physics, Queen's University, University Road, Belfast, BT7 1NN, UK\and
	Isaac Newton Group of Telescopes, Apartado de Correos 321, E-38700 Santa Cruz de la Palma, Tenerife, Spain\and
 	Astronomical Institute, Charles University Prague, V Holesovickach 2, CZ-180 00 Praha, Czech Republic\and
 	School of Physics and Astronomy, University of Manchester, M13 9PL, UK\and
	Astrophysics Research Institute, Liverpool John Moores University, CH61 4UA, UK
	}

   \date{Received August 20 2008; accepted August 20 2008}

 
  \abstract
  {
  Some of the first results are reported from RISE - a new fast camera mounted on the Liverpool Telescope primarily designed to obtain high time resolution light curves of transiting extrasolar planets for the purpose of transit timing. A full and partial transit of WASP-3 are presented, and a Monte Carlo Markov Chain analysis is used to update the parameters from the discovery paper. This results in a planetary radius of  $1.29^{+0.05}_{-0.12} R_J$ and therefore a density of $0.82^{+0.14}_{-0.09}{\rho}_J $, consistent with previous results. The inclination is $85.06^{+0.16}_{-0.15} \deg$, in agreement (but with a significant improvement in the precision) with the previously determined value. Central transit times are found to be consistent with the ephemeris given in the discovery paper. However, a new ephemeris calculated using the longer baseline results in $T_c(0) = 2454605.55915 \pm 0.00023$~HJD and $ P = 1.846835 \pm 0.000002$~days. 
}


\maketitle

\section{Introduction}

Transit observations of extrasolar planets are vital to our understanding of planetary systems, as they allow us to determine the orbital inclination and planetary radius, and when coupled with the radial velocity (RV) method, the mass and density may be derived. Ground-based transit surveys such as SuperWASP \citep{swasp} and HATnet \citep{hatnet} are now coming of age and to date over 50 transiting planets have been found\footnote{see http://exoplanet.eu/}. Most of these systems are Hot Jupiters (with the exception of GJ-436b - see \citet{gj436}), and neither these transit surveys nor the RV method are close to reaching the required precision to search for Earth-sized planets.

However, high precision transit photometry may provide us with a short-cut to detecting Earth-sized planets around existing transiting systems, as it allows us to measure the central transit times of these systems very accurately. This could reveal the presence of a third body through transit timing variations (TTV) by measuring any perturbations on the transiting planets orbital period \citep{miralda, holman_murray, agol2005, heyl_gladman}. The TTV technique is sensitive to Earth-sized planets (or smaller) when the perturber is in favourable orbital configurations - namely low-order mean motion resonances. The technology to take advantage of this technique already exists and in fact ground-based TTV experiments are the first to have probed for Earth-sized planets around main sequence stars \citep[see for example, ][]{tlc1, steffen_agol2005}.

The TTV method requires many light curves of high-time resolution coupled with precision photometry when searching for Earth-sized exoplanets. RISE (Rapid Imager to Search for Exoplanets) is a new instrument designed primarily for this purpose. It was designed and built by scientists from Queens University, Belfast, Liverpool JMU and the University of Manchester, and is mounted on the fully robotic 2-m Liverpool Telescope (LT) on La Palma. It was commissioned in late February 2008 and is described in \S2.1. 

In the months following the commissioning run some non-TTV targets were observed with RISE to test the instrument and reduction procedures further. These data have already been used to accurately characterise a recently discovered planet from the SuperWASP-N camera, WASP-14b (see \citet{wasp14} for details). In this paper we present a full transit of WASP-3 observed during this period, along with a recently obtained partial transit.  WASP-3b was also discovered by the SuperWASP-N camera \citep[see][]{wasp3} and was found to be a relatively massive planet closely orbiting an unevolved main sequence star of spectral type F7-8V.

However, there is currently only transit photometry available for WASP-3 from relatively small aperture telescopes (eg the IAC 80cm and Keele 60 cm). In this paper we analyse the RISE lightcurves to provide improved model fits of the planetary parameters and also to demonstrate the potential timing accuracy obtainable from the ground using such an instrument. We are in the process of targeting several other planetary systems with RISE in order to detect TTV signals. 

In \S2 we describe the RISE instrument and the observation and data reduction procedures. We describe the fitting technique used to extract the parameters and uncertainties from the lightcurves in \S3, and finally in \S4 and \S5 we present, discuss and summarise our results.

\section{Observations and data reduction}
\subsection{The RISE instrument}
\label{the_instrument}

RISE was mounted on the LT and commissioned in February 2008, and was designed to obtain high time precision lightcurves of transiting extrasolar planets. Due to transiting systems often being relatively bright targets ($m_v<\sim13$), overheads due to readout time adversely affect the timing precision we can obtain. Therefore, RISE consists of a thermoelectrically cooled frame transfer CCD (using an e2v CCD47-20 back illuminated chip) which allows exposure times as short as 1 second without dead time. For very bright targets (eg HD189733, $V\sim<8$) this allows much more data ($>$10000 images) to be taken during a typical 2-3 hour transit than from a traditional CCD with significant readout time. 

The CCD has $1024\times1024$ $13.0~\mu m$~pixels in the light sensitive region, giving a field of view of $9.4 \times 9.4$ arcminutes squared. This relatively large field should provide many bright sources to be used as reference stars, therefore improving the potential photometric accuracy, and allows accurate lightcurves to be taken for relatively uncrowded fields previously unattainable with RATCAM - the other imager on the LT with a field of view of 4.6 x 4.6 arcminutes squared.

Furthermore, the RISE instrument uses a single broadband filter composed of Schott KG5 and OG515, which results in a band pass of 500-700nm.
 This allows the instrument to collect as many photons as possible during an exposure, which again should improve the photometric accuracy we can obtain and therefore the timing accuracy. For more information about the design and performance of RISE, and the LT, see \citet{steele_2008}.

\subsection{RISE photometry}

A full transit of WASP-3 was observed using RISE on the night of 18 May 2008, and a partial transit was observed on the night of 4 September 2008, corresponding to epochs $E$ = 250 and 309, respectively, from the ephemeris of \citet{wasp3},
$$
T_c = 2454143.8503 + 1.846834~E~[HJD].
$$
 The data from 18 May consist of 4800 x 3s exposures beginning at 23:15 UT spanning a 4 hour period starting $\sim$0.8 hours before ingress and ending $\sim$0.4 hours after egress, resulting in an airmass range of $\sim$1.99-1.02.  The 4 September data consist of 3900 x 3s exposures beginning at 22:30 UT, spanning 3.25 hours starting approximately 0.7 hours before ingress and ending just before egress - corresponding to an airmass range of $\sim$1.08-2.10.

The detector was set to 2x2 binning mode for both transits giving a scale of 1.1 arcsec/pixel, and the LT was defocussed so that the chip was not saturated and well below the non-linearity regime resulting in a typical FWHM of $\sim$3-4 pixels (3.3-4.4 arcsec). This also helps to minimise errors associated with flat fielding. Both nights were photometric.

\begin{figure} 
\resizebox{\hsize}{!}{\includegraphics[angle=0]{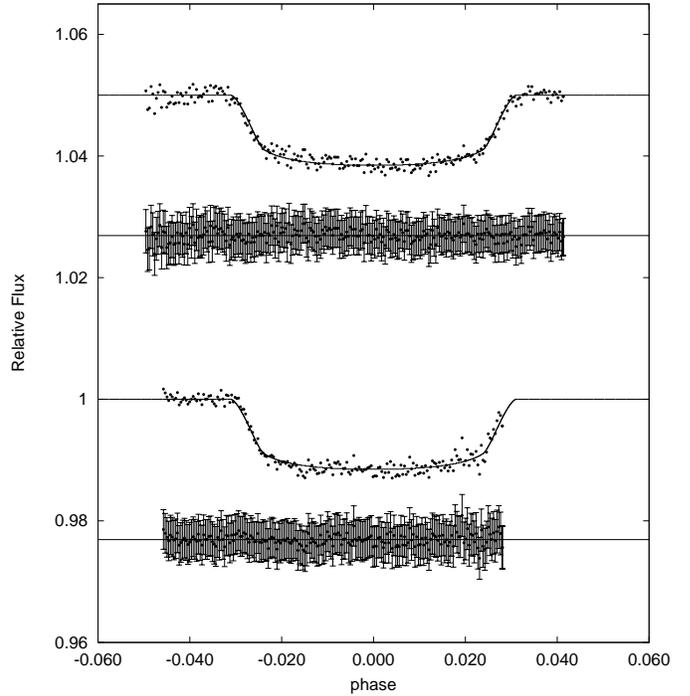}}
\caption{RISE lightcurves of WASP-3 taken on 18 May 2008 (upper, offset by 0.05) and 4 Sep 2008 (lower) shown in 1 minute bins with the best fit model from the MCMC analysis overplotted.  Residuals are shown offset along with their error bars after rescaling so $\chi^2_{red}=1$ and to account for red noise.} 
\label{fig:lightcurve} 
\end{figure}

The images were first de-biased and flat fielded with combined twilight flats using standard IRAF\footnote{IRAF is distributed by the National Optical Astronomy Observatories, which are operated by the Association of Universities for Research in Astronomy, Inc., under cooperative agreement with the National Science Foundation.} routines.  Aperture photometry was then performed on the target star and 3 nearby companion stars using Pyraf\footnote{Pyraf is a product of the Space Telescope Science Institute, which is operated by AURA for NASA.} and the DAOPHOT package using a 7 pixel aperture (the same companion stars were not used in both cases due to different field orientations). The flux of WASP-3 was then divided by the sum of the companion stars (all checked to be non-variable) to obtain each light curve. Initial estimates of the photometric errors were calculated using the aperture electron flux, sky and read-noise. The lightcurves were then normalised by dividing through with a linear function of time fitted to the out-of-transit data, setting the unocculted flux of WASP-3 equal to 1, revealing a transit of depth $\sim$1.17\%. The lightcurves along with their best fit models and residuals are shown in Fig.~\ref{fig:lightcurve}, after grouping the data into 1 minute bins.  The rms of the residuals from the best fit model (see \S3.1) are $\sim4.3$~mmag for each lightcurve per 3s exposure.

\section{Results and analysis}

\subsection{Determination of system parameters}

In order to determine the system parameters from the photometry, a parameterised model was constructed. This used Kepler's Laws and assumed a circular orbit to calculate the normalised separation ($z$) of the planet and star centres as a function of time from the stellar mass and radius  ($M_{\star}$ and $R_{\star}$), the planetary mass and radius  ($M_{p}$ and $R_{p}$), the orbital period and inclination ($P$ and $i$), and finally a central transit time for each lightcurve ($T_{0,n}$).  The analytic models of \citet{mandel_agol} were then used to calculate the stellar flux occulted by the planet from the normalised separation and the planet/star radius ratio ($\rho$) assuming the quadratic limb darkening function
$$
\frac{I_{\mu}}{I} = 1 - a(1-\mu) - b(1-\mu)^2,
$$
where $I$ is the intensity, $\mu$ is the cosine of the angle between the line of sight and the normal to the stellar surface, and $a$ and $b$ are the linear and quadratic limb darkening coefficients, respectively.

The limb darkening parameters were obtained from the models of \cite{claret}. We linearly interpolated the ATLAS tables for $T_{eff}$ = 6400 K, log $g$ = 4.25, [Fe/H] = 0.0 and $v_{t}$ = 1.0 km s$^{-1}$ \citep[from][]{wasp3} to obtain limb darkening parameters in both the V and R bands. The average of the parameters from each of the bands was then taken to be our theoretical limb darkening parameters.

A Markov Chain Monte Carlo (MCMC) algorithm was then used to obtain the best fit parameters and their uncertainties \citep[see for example, ][]{tegmark, tlc1, cameron, tlc9}. This consists of calculating the $\chi^2$ fitting statistic,
$$
\chi^2=\sum_{j=1}^{N}\frac{( f_{j,obs} - f_{j,calc})^2}{\sigma^2_j},
$$
where  $f_{j,obs}$ is the flux observed at time $j$, $\sigma_j$ is the corresponding uncertainty and  $f_{j,calc}$ is the flux calculated from the model for time $j$ and for the set of physical parameters described above. The best fit parameters and uncertainties are then found by starting a long chain that explores the probability densities in parameter space by accepting or rejecting subsequent parameter sets based on the difference in $\chi^2$.  The new parameter set is accepted if $\chi^2$ is lower than that of the previous set, or accepted with probability 
$$
P = \exp(\frac{-\Delta\chi^2}{2})
$$
if $\chi^2$ is higher than that of the previous parameter set. New parameter sets are chosen by adding a small random amount to each of the previously accepted parameters determined by a jump function. These jump functions consist of a random value selected from a Gaussian distribution with mean 0 and standard deviation of 1, scaled by a factor associated with each parameter (which ideally should initially be set near to each parameter's 1$\sigma$ uncertainty to speed chain convergence). These scale factors were then resized from their initial values so that $\sim$25\% of all new parameter sets were accepted. The reader is referred to the aforementioned literature for more details of MCMC fitting techniques.

To obtain reliable estimates of parameters and their uncertainties it is important that the photometric errors are calculated accurately. The photometric errors $\sigma_j$ are first rescaled so that the best fitting model for each lightcurve has a reduced $\chi^2$ ($\chi^2_{red,n}$) of 1, which required the initial error estimates to be multiplied by the factors $\sqrt{\chi^2_{red,1}} =$ 1.73 and$\sqrt{\chi^2_{red,2}} =$ 2.04. This indicates that the photometric errors are considerably higher than what would be expected from photon noise alone.

It is also vital to account for any correlated ('red') noise in the data \citep[see][]{pont2006, gillon2006}. Following \citet{tlc9} we evaluated the presence of red noise in each lightcurve by calculating a factor $\beta$ and again rescaling the $\sigma_j$ in the $\chi^2$ calculation by this value. A value for $\beta$ is determined by analysing the residuals from the best fit model of the lightcurves. Calcualting the standard deviation of the residuals $\sigma_1$ and also the standard deviation after binning the residuals into $M$ bins of $N$ points $\sigma_N$, one would expect
$$
\sigma_N = \frac{\sigma_1}{\sqrt{N}}\sqrt{\frac{M}{M - 1}}
$$
in the absence of red noise. However, this is usually larger by a factor $\beta$. As pointed out in \citet{tlc9}, $\beta$ only depends weakly on the averaging time, therefore we used the average of $\beta$ values calculated from a range of averaging times from 10-30 min to rescale the photometric errors. The values of $\beta$ used were 1.43 and 1.35 for the full transit and partial transit, respectively.

To account for any normalisation errors in the lightcurves, a further 2 free parameters were added for each transit.  These were the out-of-transit flux ($f_{oot,n}$) and a time gradient ($t_{Grad,n}$), which are vital for TTV measurements as these effect the symmetry of the lightcurve and therefore the central transit times. An airmass correction $k$ was tried for the full transit instead of the time gradient by multiplying each point by a factor $exp(-kz)$ where $z$ is the airmass. This was found to cause an insignificant change in the output parameters and so the time gradient was used in our final analysis. The chains would not converge when an airmass correction was used instead of a time gradient for the partial transit.

A Gaussian prior was added to the $\chi^2$ fitting statistic so that
$$
\chi^2=\sum_{j=1}^{N}\frac{( f_{j,obs} - f_{j,calc})^2}{\sigma^2_j}  
+ \frac{(M_{\star,j} - M_0)^2}{{\sigma_M}^2},
$$
where $M_{\star,j}$ is the mass of the star used to calculate each model at time $j$, and $M_0$ and $\sigma_M$ are the stellar mass and corresponding uncertainty as given in \citet{wasp3}. This is so the error in the stellar mass is taken into account when determining parameters from the MCMC probability distributions. The stellar radius cannot be allowed to vary independently of the stellar mass as the chains will not converge. Therefore, the stellar radius was updated for each choice of $M_{\star}$ using the scaling relation $R_{\star} \propto M_{\star}^{1/3}$, whilst $P$ and $M_{p}$ were held fixed at their previously determined values, as their uncertainties do not have any significant effect on the output probability distributions. 

An initial MCMC analysis was used to estimate the initial parameters and jump functions for $\rho$, $i$, $T_{0,n}$, $f_{oot,n}$ and $t_{Grad,n}$. Five separate chains with 100000 points were then computed with the initial free parameters set by adding a $5\sigma$ gaussian random to their previously determined best fit values.  The first 20\% of each chain was eliminated to keep the initial conditions from influencing the results, and the remaining part of the chains were merged to obtain the best fit parameters and uncertainties. These were not assumed to be Gaussian and were obtained by setting the best fit value to the modal value of the probability distribution, and the $1\sigma$ lower (upper) limit as the value where the integral of the distribution from the minimum (maximum) value to it was equal to 0.159. To test that the chains had all converged to the same parameter space, the Gelman \& Rubin statistic was then calculated for each of the free parameters (this time assuming Gaussian errors) \citep{gelrub}. It was found to be less than 0.5\% from unity for all parameters, a good sign of mixing and convergence. The transit duration ($T_d$) and impact parameter ($b$) were also calculated for each parameter set in the chain using the equations of \citet{seager2008}, and were treated the same as the fitted parameters to determine best fit values and uncertainties. This was to account for any correlated errors.

To check for any errors that may have resulted from a poor choice of limb darkening parameters, the above procedure was repeated, this time allowing the linear limb darkening parameter ($a$) to vary freely whilst holding the quadratic limb darkening parameter ($b$) fixed at the theoretical value obtained from the models of \citet{claret}. This is the same procedure used by \citet{tlc9}, and did not cause any significant changes to the best fit values, however, it did increase the errors in $i$ and $\rho$ slightly and hence these results were adopted as our final values. A further check on the choice of limb darkening parameters involved repeating this process by replacing the quadratic limb darkening coefficient ($b$) determined for the combined V+R filter by those obtained for the individual V and R filters. This caused no significant variation from the previously determined results in either case, and thus we conclude that the choice of limb darkening parameters had no detrimental effects on our results. This was not surprising, as the linear limb darkening parameter was much more sensitive to the choice of filter than the quadratic parameter.

\section{Results}

\begin{table*}
\caption[]{Parameters and $1\sigma$ uncertainties for WASP-3 as derived from MCMC fitting and some further calculated parameters. The planetary mass was taken from the discovery paper and is displayed for convenience.}
\label{tab:parameters}
\begin{center}
\begin{tabular}{lccc}
 Parameter & Symbol & Value & Units\\
 \hline\\
Planet/Star radius ratio	&	$\rho$         	& $0.1014^{+0.0010}_{-0.0008}$ 				&\\
Orbital inclination 		&	$i$  			& $85.06^{+0.16}_{-0.15}$		     			& $\deg$\\
Impact parameter		&	$b$  			& $0.448^{+0.014}_{-0.014}$ 					&\\ 
\\
Transit duration		&	$T_d$		& $2.753^{+0.020}_{-0.013}$		 			& hours\\
Transit epoch			&	$T_0$		& $2454605.55915 \pm 0.00023$ 				&HJD\\
Period				&	$P$			& $1.846835 \pm 0.000002$					& days\\
\\
Planet radius			&	$R_p$         	& $1.29^{+0.05}_{-0.12}$ 						& $R_J$\\
Planet mass			&	$M_p$  		& $1.76^{+ 0.08}_{-0.14}$  					& $M_J$\\ 
Planet density			&	${\rho}_p$  	& $0.82^{+0.14}_{-0.09}$						& ${\rho}_J$\\ 
Planetary surface gravity	& 	log $g_p$		& $3.42^{+0.06}_{-0.04}$                                             & [cgs]\\

\\

\hline\\
\end{tabular}
\end{center}

\end{table*}

\subsection{MCMC results}

The parameters and their uncertainties calculated from the MCMC fit are shown in Table~\ref{tab:parameters}.  The planet radius was found to be $1.29^{+0.05}_{-0.12}~R_J$, and the planet density $0.82^{+0.14}_{-0.09}~\rho_J$ - consistent with that of the discovery paper. Therefore the derived planetary parameters are still in agreement with models of highly irradiated Hot Jupiters \citep[e.g.][]{fortney_2007} as concluded in \citet{wasp3}.

The inclination is also consistent with that calculated previously but it is now known much more accurately with a value of $85.06^{+0.16}_{-0.15}\deg$ - a factor of $\sim$5 increase in precision. However, the transit duration was found to be inconsistent with that of the discovery paper $(\sim4.1\sigma)$. $T_d$ is very sensitive to the inclination of the system, and we believe the discrepancy arises due to the discovery data not being of sufficient quality to constrain it accurately, and therefore has resulted in an underestimated error in the transit duration. If the updated inclination from the RISE data is used along with the remaining parameters from \citet{wasp3}, then the calculated $T_d$ is found to be consistent with our value.

\subsection{Transit ephemeris}

\begin{figure} 
\resizebox{\hsize}{!}{\includegraphics[angle=0]{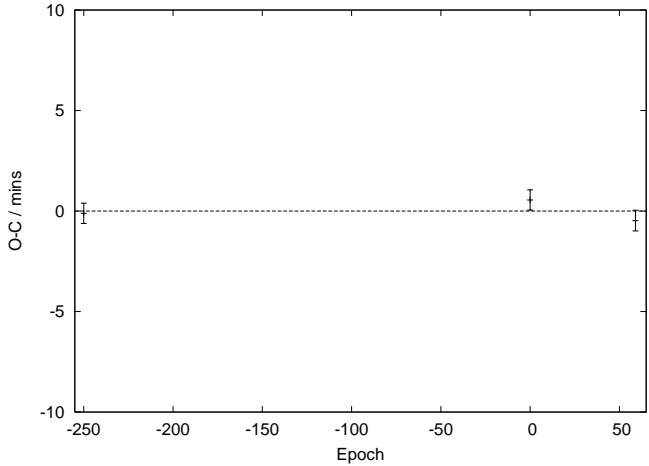}}
\caption{Timing residuals of the RISE transits of WASP-3 and the epoch given in \cite{wasp3} using the updated ephemeris.} 
\label{fig:ttvplot} 
\end{figure}

The RISE transits do not differ significantly from the discovery ephemeris however this can be updated using the longer baseline. This was found by minimizing $\chi^2$ by fitting a linear function of Epoch $E$ and period $P$ to the transit times from this paper (shown in table~\ref{tab:timings}) and the ephemeris of \cite{wasp3},
$$
T_c(E) = T_c(0) + EP,
$$
where $T_c(0)$ was set to the central transit time from 18 May 2008 RISE data. The results were $T_c(0) = 2454605.55915 \pm 0.00023$~[HJD] and 
$P = 1.846835 \pm 0.000002$~days. Fig~\ref{fig:ttvplot} shows a plot of the timing residuals of WASP-3 transits using the updated ephemeris.

\begin{table}
\caption[]{Central transit times and uncertainties for the RISE photometry.}
\label{tab:timings}
\begin{center}
\begin{tabular}{lcc}
Epoch & Transit Time [HJD] & Uncertaintly (days)\\
\hline\\
0 & 2454605.55956 & 0.00035   \\
59 & 2454714.52210 & 0.00036 \\ \\
\hline
\end{tabular}
\end{center}
\end{table}

\section{Summary and Discussion}

We have presented some of the first results from RISE of a full and partial transit of the exoplanet system WASP-3. An MCMC analysis was used to update the parameters which were found to be consistent with those of \citet{wasp3}, except for the calculated value for the transit duration, which is explained in terms of the large uncertainty in the previously determined inclination. Using the central transit times an updated ephemeris was calculated to aid in further observations of this system.

Although the timing accuracy was not of the order we intended for RISE ($<\sim10s$), other transits observed with RISE (for TTV experiments) are producing much more impressive light curves \citep[see][]{steele_2008}. This is mainly because the photometric errors for each WASP-3 light curve had to be multiplied by the factors $\sqrt{\chi^2_{red}} = $ 1.73 and 2.04 in our analysis so that the reduced $\chi^2 = 1$.  It is as yet unclear why this is the case, but is perhaps due to randomised systematics caused by a combination of the moon, changing flat-field structure and poor autoguiding. Further RISE observations currently being made at the LT should not suffer from these problems, and we hope to report on further transit timing datasets in future papers.

\begin{acknowledgements}
RISE was designed and built with resources made available from Queens University Belfast, Liverpool John Moores University and the University of Manchester. The Liverpool Telescope is operated on the island of La Palma by Liverpool John Moores University in the Spanish Observatorio del Roque de los Muchachos of the Instituto de Astrofisica de Canarias with financial support from the UK Science and Technology Facilities Council. F.P.K. is grateful to AWE Aldermaston for the award of a William Penney Fellowship. We extend our thanks to members of the Liverpool Telescope support staff, who have been invaluable during the commissioning period of RISE and subsequent test observations. We finally thank the anonymous referee for helpful comments that led to several improvements in the manuscript.
\end{acknowledgements}

\bibliographystyle{aa} 
\bibliography{mybib} 

\end{document}